# Generation of Ultrafast Jets through the Implosion of a Nested Conical Wire Array

F. Winterberg

University of Nevada, Reno

**Abstract**

It is shown that in contrast to the electric pulse power driven implosion of a single conical wire array, the implosion of a nested conical wire array with opposite alternate opening angles can lead to the generation of fast jets, with velocities of the order $10^8$ cm/s. This technique can be applied for the supersonic shear flow stabilization of a dense z-pinch, but possibly also for the fast ignition of a pre-compressed dense deuterium-tritium target.

1. **Introduction**

Back in 1967 the author had proposed to reach very large jet velocities by the impact under a small angle of a projectile on a stationary solid target [1]. For impact velocities of ~$10^7$ cm/s and an angle of less than 10 degrees, jet velocities of the order $10^8$ cm/s could be expected, as they are needed for impact fusion [2]. Projectile velocities of ~$10^7$ cm/s can in principle be reached by the acceleration of a small superconducting solenoid with a magnetic travelling wave accelerator, but the length of such an accelerator was estimated to be of the order 10 km [3, 4]. However, it has been shown that such velocities can also be reached over a distance of a few cm by the electric pulse power driven implosion of a cylindrical thin wire array [5, 6]. This raises the question if with this technique jet velocities of the order $10^8$ cm/s can be reached by the implosion of a conical wire array with a small opening angle. It turns out that this is not possible, but possible with a nested conical wire array.

2. **Some remarks about the fundamental importance to reach large jet velocities by oblique implosion geometries.**

In the geometry of reference 1 the jet is planar. Higher jet energy densities can be expected by going from a planar to a conical geometry. In either case the physics is the same. As illustrated in Fig.1 the imploding cone produces a "thin" collimated jet moving with a high velocity to the right, with its momentum balanced by the momentum of a "fat slug", moving to the left with a much smaller velocity.

If the cone implodes with a velocity $V_0$ in a direction perpendicular to the surface of the cone, and if ½ of the opening angle of the cone is α, the velocity $v_j$ of the jet and $V_s$ the slug are [7]:

$$v_j = (V_0/\sin\alpha)(1+\cos\alpha) \tag{1a}$$

$$V_s = (V_0 / \sin\alpha)(1 - \cos\alpha) \tag{1b}$$

For α <<1, approximately:

$$v_j \simeq 2V_0 / \alpha \tag{2a}$$

$$V_s \simeq V_0 \alpha / 2 \tag{2b}$$

This shows that with α ⟶ 0, $v_j$ ⟶ ∞ and $V_s$ ⟶ 0.

If $M_0$ is the total mass of the cone

$$M_0 = M_s + m_j \tag{3}$$

where $M_s$ and $m_j$ are the mass of the slug and the jet, one has

$$M_0 = (M_0 / 2)(1 + \cos\alpha) \tag{4a}$$

$$m_j = (M_0 / 2)(1 - \cos\alpha) \tag{4b}$$

for α<<1,

$$M_s \simeq M_0 \tag{5a}$$

$$m_j \simeq M_0 \alpha^2 / 4 \tag{5b}$$

This shows that with α ⟶ 0, $M_s$ ⟶ $M_0$ and $m_j$ ⟶ 0.

From (1a), (1b), (4a) and (4b) one also obtains

$$m_j \simeq M_0 \alpha^2 / 4 \tag{6}$$

and that the momentum of the slug is equal and opposite to the momentum of the jet.

One also finds for the total kinetic energy

$$(1/2)M_0 V_0^2 = (1/2)(M_s V_s^2 + m_j v_j^2) \tag{7}$$

This result is to be expected, because the flow of the imploding cone is isentropic. In reality some of the energy is lost primarily by black body radiation. However, if the cone implosion is driven by electric pulse power, the lost energy can simply be added by increasing the input of the electromagnetic energy.

Of interest is also the efficiency η of the conversion of the energy (1/2) $M_0 V_0^2$ into the energy of the jet (1/2) $m_j v_j^2$, one obtains:

$$\eta = \frac{(1/2)m_j v_j^2}{(1/2)M_0 V_0^2} = \frac{1}{2}(1+\cos\alpha) \tag{8}$$

It follows that for α ⟶ 0, η ⟶ 1, favoring small angles α.

The conical implosion solution is an exact solution of the Euler equation conserving energy and momentum.

### 3. Implosion of a conical wire array

The large velocities obtained in the implosion of cylindrical wire arrays raises the question if the described kind of jet formation might be possible with the implosion of conical wire arrays, reaching with small opening angles even much higher velocities. It is easy to show that this is not possible. For this to happen the sequence of the events during the implosion must have a pattern as shown in Fig. 2a, requiring that the Alfven velocity $v_A = B/\sqrt{4\pi\rho}$ (B magnetic field strength, ρ density), which has to be set equal the implosion velocity $V_0$, must be constant along the wires. This is certainly not the case as can be seen as follows: The magnetic field B, averaged over a circle of the radius r measures from the axis of the cone, goes as B= const./r. and the likewise averaged density ρ of the wire array also goes as 1/r. Accordingly, the Alfvén velocity goes as $1/\sqrt{r}$. As a result, the discharge over a conical wire array will not produce a jet but rather a "blob" as shown in Fig. 2b, moving to the right with the velocity of the vertex point of the imploded wire array.

### 4. Motivation

The desire for the generation of a fast jet arose from the idea that the long sought solution of the problem to stabilize the linear pinch discharge could be by the superposition of a supersonic flow over or through the pinch discharge channel [8, 9]. The flow must be supersonic, because only then does the stagnation pressure of the flow overwhelm the magnetic pressure of the pinch discharge. At thermonuclear temperatures of ~$10^8$ K, where the thermal ion particle velocity is ~$10^8$ cm/s, supersonic means plasma velocities in excess of ~$10^8$ cm/s. For a Mach number M equal to M≈3, a temperature of ~$10^9$ K would be needed if these velocities are generated by thermal expansion. It is for this reason that the author had suggested to reach these velocities with a high current particle accelerator [10, 11], but this requires a large effort. To circumvent this problem it was proposed to shoot a thin needle-like metallic projectile through the core of the pinch discharge channel [8]. There, because of the high density and low temperature of the metallic projectile, the attainment of supersonic velocities requires only a velocity of 10 km/s, which could be reached with a light gas gun. For a projectile density $\rho \approx 10$ g/cm$^3$, and a velocity of $v \approx 10^6$ cm/s, the stagnation pressure is $\rho v^2 \sim 10^{15}$ dyn/cm$^2$. If set equal to the magnetic pressure of the pinch discharge channel $H^2/8\pi$, one finds that $H \sim 10^7$ G. The pinch current to reach this magnetic field is $I \sim 5 \times 10^7$ r, where r is the pinch radius. For $r \sim 1$ cm, the current would be $I \sim 5 \times 10^6$ A, still large enough to entrap most of the charged fusion product particles (α-particles), of the deuterium-tritium reaction inside the pinch channel. One important benefit of this idea is that through the temperature gradient between the cool metallic projectile and the hot plasma surrounding the projectile, large currents are induced by the thermomagnetic Nernst effect in the boundary layer between both. These currents lead to a strong magnetic field repelling, and thereby thermally insulating, the hot plasma against the much cooler projectile.

A somewhat similar proposal was made by Hassan and Yi-Min Huang [12], where a fast plasma jet is shot into a metallic torus. The Nernst effect creates there a large magnetic field between the hot plasma and the cold wall of the torus, thermally insulating the plasma against the wall, and in addition stabilizing the plasma by its shear flow created in the boundary layer along the wall.

One problem with the supersonic shear flow stabilization of a linear pinch discharge is that the axial flow removes a large amount of heat, lowering the energy confinement time. This problem does not arise in the toroidal configuration of Hassam and Yi-Min Huang, but it likely will make unfeasible the linear supersonic shear flow pinch stabilization concept by Arber and Howell [9], and by Shumlak et al. [13].

In a proposal put forward by the author many years ago [14], this does not happen, because there the pinch current is assumed to be about $10^7$ A, or larger, where the charged DT fusion reaction products are entrapped inside the pinch discharge channel, making possible the ignition of a linear thermonuclear detonation wave propagating down the channel with the velocity of $\sim 10^9$ cm/s. Because the pinch must there have the highest possible density, realized at its lowest possible temperature, the supersonic flow velocity, required to stabilize the pinch discharge, can there be much smaller. And it was proposed to generate a supersonic jet flowing over the pinch discharge by the implosion of a cylindrical wire array, where the soft X-rays emitted from the imploded wire array ablatively drive the jet [15]. An alternative was the idea to use the implosion of a conical wire array [16]. But since this configuration does not lead to a jet but rather to a "blob", it was in an unpublished note proposed by the author that a jet could there be produced by making the wires thinner along their length. This proposal was difficult to realize and was never carried out. In the following section a different solution is presented.

## 5. Solution with two nested wire arrays

The solution to prevent the formation of a blob is to use nested wire arrays. It can already be implemented by two nested wire arrays. This configuration is shown in Fig. 3.

The inner wire array, II, forms a divergent cone from the point A where in a cylindrical coordinate system r = z = 0, at one side of a planar high voltage diode. It ends at the circle C at the other side of the diode where r = $r_1$, z = $\ell$, with $\ell$ the width of the diode. Unlike the inner wire array, II, the outer wire array, I, has the form of a cut-off cone, stretching from the position B, where r=$r_0$, z=0 at the left side of the diode, to the position C on the right side of the diode where r = $r_1$, z = $\ell$, touching at that position the inner wire array.

If a high voltage pulse is applied to the diode, a current flows from B to C, because the impedance of the inner wire array is much larger, and no current flows through the inner wire array.

The current flowing through the outer wire array implodes the outer array with the Alfven velocity $v_A = B/\sqrt{4\pi\rho}$, from the position r = $r_0$, z=0; With B ∝ 1/r, ρ ∝ 1/r, with the velocity

$$v_0 = -a/\sqrt{r} \text{ , a = const.} \tag{9}$$

From the position r = $r_1$, z = $\ell$, the imploding outer wire array entrains the inner wire array, increasing its density two-fold. The implosion velocity from the position r = $r_1$, z = $\ell$ towards r = 0, z = $\ell$, thus is

$$v_1 = -a/\sqrt{2r} \tag{10}$$

This two-fold implosion process results in a narrow cone with a small opening angle, as it is required for a large jet velocity. From a small opening angle, the time $t_0$ for the implosion of the outer array in going from r = $r_0$, z = 0, to r = z = 0, should only be slightly shorter that the time $t_1$ for the implosion to go from the position r = $r_1$, z = $\ell$ to r = 0, z = $\ell$.

One obtains

$$t_0 = -\frac{1}{a}\int_{r_0}^{0}\sqrt{r}\,dr = \frac{2}{3a}r_0^{3/2} \tag{11a}$$

$$t_1 = -\frac{\sqrt{2}}{a}\int_{r_1}^{0}\sqrt{r}\,dr = \frac{2^{3/2}}{3a}r_1^{3/2} \tag{11b}$$

To make $t_0 < t_1$ then requires that

$$r_0 \leq 2^{1/3} r_1 \tag{12}$$

In the limit where $r_0 = 2^{1/3} r_1$, the opening angle of the implosion generated cone would equal to zero.

If for $r_1 = 2^{-1/3} r_0 < r_0$, the angle would be zero, it would be non-zero for $r_1 > 2^{-1/3} r_0$, and it would be also non-zero for $r_1 = r_0$. To compute the angle for $r_1 = r_0$, we first compute the time $t_1$ for the implosion at position C to reach from $r_1 = r_0$ at C the distance $r = \Delta r$ from the z-axis:

$$t_1 = -\frac{\sqrt{2}}{a}\int_{r_0}^{\Delta r}\sqrt{r}\,dr = \frac{2^{3/2}}{3a}\left[r_0^{3/2} - (\Delta r)^{3/2}\right] \tag{13}$$

Setting $t_1$ equal to $t_0$ for the time from position B to reach $r = 0$, given by (11a), we find that

$$\frac{\Delta r}{r_0} = \left(1 - \frac{1}{\sqrt{2}}\right)^{2/3} \simeq 6.6 \times 10^{-2} \tag{14}$$

Assuming that $\ell = 2r_0$, one finds $\Delta r/l \sim 10^{-1} \sim \alpha$. If according to (2a) $V_0 \sim 10^7$ cm/s, then $v_j \sim 2 \times 10^8$ cm/s.

## 6. Application to fast ignition and magnetized target fusion

We briefly discuss here two applications to nuclear fusion:

1. <u>Fast ignition of a highly compressed deuterium-tritium(DT) target:</u>

The idea to ignite the target with the jet from a laser induced implosion of a cone attached to the DT target was proposed by Velarde et al. [17, 18]. These authors were apparently unaware of the 1967 proposal by the author [1], where instead of a laser beam a fast moving projectile imploded the cone. Regarding the proposal by Velarde et al., it was pointed out by Murakami and Nagatomo [19], that because of the proximity of the jet-producing cone to the DT target, "the specific momentum of the jet, $\rho v^2$, is at best in the same order of that of the main fuel". But if the jet is produced by electric pulse power, and if its origin is spatially separated from the DT target, the energy density of the jet can be larger. And with electric pulse power cheap in comparison to laser pulse power, small energy conversion efficiency into jet energy can here be tolerated. Moreover, experiments made with conical liners imploded by high explosives have shown that for small opening angles and the highest jet velocities, the jet has a tendency to condense into dust particles [20]. This behavior has also been observed in the supersonic expansion flow of hydrogen [21]. Such a condensation, of course, is possible only if the jet has enough time to cool down to a low temperature, not possible for a jet which is produced in the immediate vicinity of the DT target. It is then conceivable that the condensation may produce the fast moving small solid projectile needed in the fast impact ignition concept by Murakami and Nagatomo [19]. The target would have here to be pre-compressed just in the moment the projectile strikes the target.

2. <u>Supersonic flow stabilization magnetized dense target:</u>

Here we consider a modified version of the idea proposed by the author to stabilize a z-pinch by shooting through its core a thin rod [8]. There it was required that the stagnation pressure $(1/2)\rho v^2$ of the fast moving rod exceeds the magnetic pressure $H^2/8\pi$ of the pinch discharge. With $\rho \sim 10$ g/cm$^3$, $v \sim 10^6$ cm/s (attainable with a light gas gun), one has $(1/2)\rho v^2 \sim 5 \times 10^{12}$ dyn/cm$^3$, and $H \leq 10^7$ G. In this example the rod moves supersonically, because the velocity of sound in it is about $10^5$ cm/s. One can then say that the fast moving rod stabilizes the pinch discharge by the shear flow set up in the boundary layer between the rod and the pinch.

In the proposed modification, the rod is replaced by the much faster moving jet which is shot into liquid DT placed into an externally applied axial magnetic field. At a jet velocity of $\sim 10^8$ cm/s, the boundary layer between the jet and DT has a temperature of $\sim 10^8$ K, which is the ignition temperature of DT, eliminating the need for an auxiliary pinch discharge. And because of the thermomagnetic Nernst effect, an azimuthal current is set up around the jet, with the externally applied magnetic field acting as a seed-field for this thermomagnetic dynamo [8, 12].

We can illustrate this idea with an example:

Liquid DT with $n = 5 \times 10^{22}$ cm$^{-3}$ heated to $T \sim 10^8$ K, has a pressure $p = 2nkT \sim 10^{15}$ dyn/cm$^2$. This pressure must be overcome by stagnation pressure of the jet $p_s = (1/2)\rho_j v_j^2$, where $\rho_j$ is the density of the jet. Assuming that $v_j \sim 10^8$ cm/s, one finds that to make $p_s > p$, one must have $\rho_j > 0.2$ g/cm$^3$. If the jet is made from the dense material of the imploding wire array with ($\rho$

~ 20 g/cm$^3$ for tungsten), it follows that the needed jet density must be not less than (1/100) of solid density.

Further, to equate the jet stagnation pressure $p_s \sim 10^{15}$ dyn/cm$^2$, with the magnetic pressure $H^2/8\pi$, one finds that $H \sim 10^8$ G, produced by the Nernst effect.

In the Lawson criterion

$$n\tau > 10^{14} s/cm^3 \tag{15}$$

We have to set $n = 5 \times 10^{22}$ cm$^{-3}$, and $\tau \sim \ell/v$, where $v \sim 10^8$ cm/s and $\ell$ the length of the jet. We find that $\ell > 0.2$ cm. With the condition for propagating detonation burn of a magnetized target given by [22]

$$Hl > 10^7 Gcm \tag{16}$$

this condition is well satisfied for a jet with a length $\ell > 1$ cm.

### 7. Conclusion

It is shown that the generation of the ultrafast jets can be generated by nested conical wire arrays, not possible with single conical wire arrays. In this concept the direction of the opening angle between consecutive nested wire arrays is alternating. The attractiveness of this proposed technique is its simplicity and it can already implemented with two wire arrays.

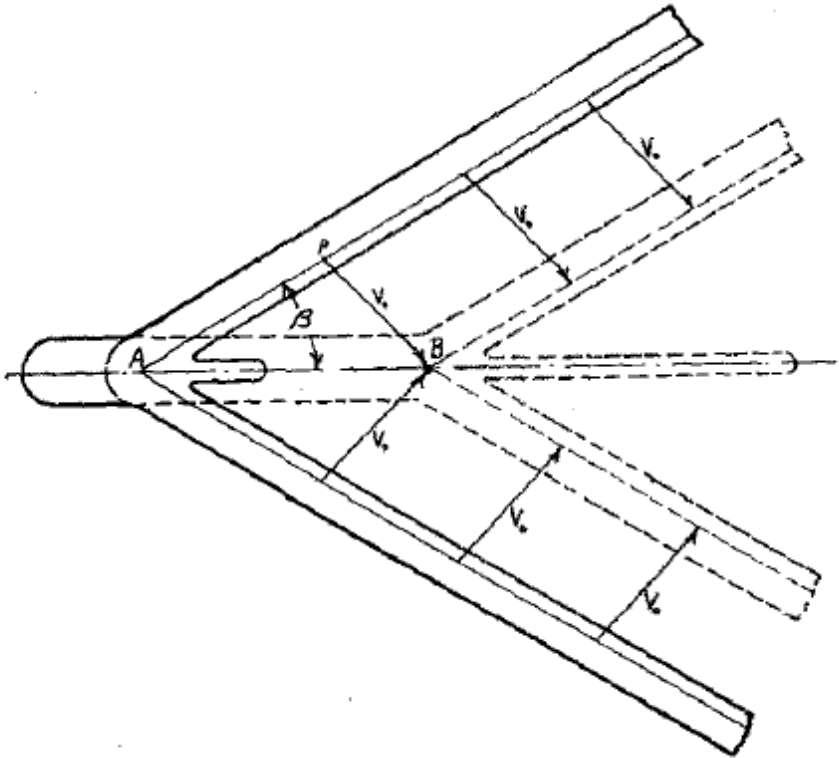

**Fig.1.** Formation of jet and slug from a cone or wedge shaped liner whose sides collapse with constant velocity $v_0$ as a result of the explosion of a charge that was in contact with the outer surface. The solid lines show conditions at an early instant of time, and the dotted lines show conditions after the walls have moved a distance equal to the velocity $v_0$ [7].

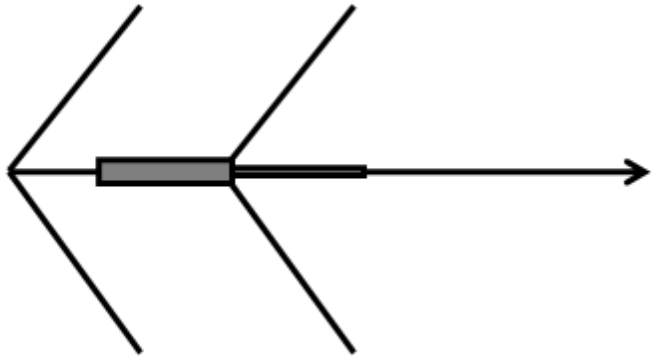

Fig. 2a

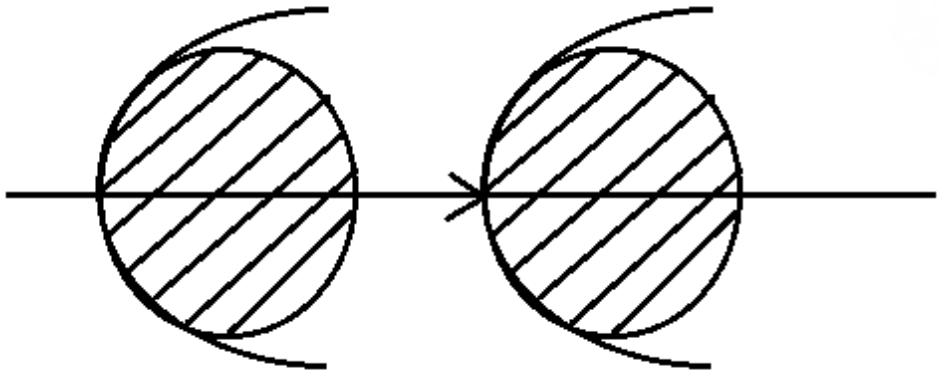

Fig. 2b

**Fig.2. Jet versus "blob" in different conical implosion configurations.**

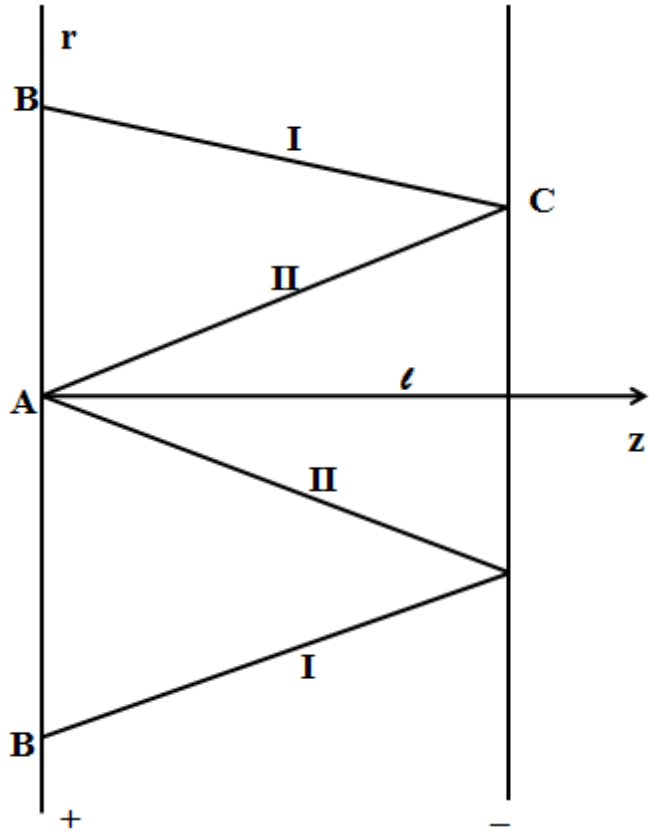

**Fig.3. Double nested coaxial conical wire array: I-Outer convergent conical array, II-Inner divergent conical array.**